\documentclass[aip,
 prd,
 jmp,
 preprint,
 reprint,
 twocolumn,
 superscriptaddress,
 author-year,
 author-numerical,
 ,10pt]{revtex4-2}
\usepackage{tikz}
\usepackage{tikz-feynman}
\usetikzlibrary{decorations.pathmorphing,decorations.markings}
\usepackage{amssymb}
\usepackage{pgfplots}
\usepackage{amsmath}
\usepackage{graphicx}
\usepackage{dcolumn}
\usepackage{bm}
\usepackage{xcolor}
\usepackage{float}
\usepackage{subfig}
\usepackage{comment}
\usepackage{multirow}
\usepackage[figurename=FIGURE,tablename=TABLE]{caption}
\usepackage{threeparttable}
\usepackage{hyperref}
\hypersetup{
    colorlinks=true, 
    linktoc=all,     
    linkcolor=magenta,
    citecolor=red
}

\begin{document}

\title{Wald Entropy in Extended Modified Myrzakulov Gravity Theories: \(f(R, T, Q, R_{\mu\nu}T^{\mu\nu}, R_{\mu\nu}Q^{\mu\nu}, \dots)\)}

\author{Davood Momeni}
    \affiliation{Department of Physics \& Pre-Engineering , Northeast Community College, 801 E Benjamin Ave Norfolk, NE 68701, USA}

\author{Ratbay Myrzakulov}
\affiliation{Ratbay Myrzakulov Eurasian International Centre for Theoretical Physics, Astana 010009, Kazakhstan}
\affiliation{L. N. Gumilyov Eurasian National University, Astana 010008, Kazakhstan}

\date{\today}

\begin{abstract}
We investigate black hole entropy in a broad class of modified Myrzakulov gravity theories defined by generalized Lagrangians of the form \( \mathcal{L} = \alpha R + F(T, Q, R_{\mu\nu}T^{\mu\nu}, R_{\mu\nu}Q^{\mu\nu}, \dots) \), where \( R \), \( T \), and \( Q \) represent curvature, torsion, and non-metricity scalars. Using the vielbein formalism, we derive the Wald entropy for various subclasses of these models, extending the classical entropy formula to accommodate non-Riemannian geometry. Our focus is on how the additional geometric degrees of freedom modify the entropy expression. The analysis shows that such corrections arise systematically from the extended structure of the action and preserve diffeomorphism invariance. These results refine the theoretical framework for gravitational thermodynamics in extended geometry settings.
\end{abstract}

\maketitle
\tableofcontents
\newpage

\section{Motivation for Modified Gravity Theories}

Einstein's General Relativity (GR), formulated in 1915, remains an elegant and experimentally successful theory of gravitation. Nonetheless, several observational and theoretical issues have motivated the exploration of its extensions. Among these are the discovery of cosmic acceleration \cite{Riess1998, Perlmutter1999}, the unknown nature of dark energy and dark matter \cite{Copeland2006}, the need for a mechanism driving inflation \cite{Starobinsky1980}, and the absence of a consistent quantum theory of gravity.

Data from the cosmic microwave background (CMB) \cite{Planck2018}, baryon acoustic oscillations (BAO) \cite{Eisenstein2005}, and supernovae \cite{Riess1998, Perlmutter1999} suggest the accelerated expansion of the universe, prompting modifications of the Einstein–Hilbert action. These extensions often invoke Myrzakulov generalized geometric scalars and preserve second-order field equations:
\begin{itemize}
    \item $f(R)$ gravity modifies curvature-based dynamics \cite{Sotiriou2010},
    \item $f(T)$ gravity employs torsion in a teleparallel formulation \cite{Cai2016},
    \item $f(Q)$ gravity incorporates non-metricity \cite{Jimenez2018}.
\end{itemize}

Metric-affine gravity generalizes these approaches by allowing the metric and affine connection to vary independently, enabling non-minimal couplings of $R$, $T$, and $Q$ \cite{BeltranJimenez2020}. Noether symmetry techniques guide model construction \cite{Capozziello1996, Paliathanasis2014}, while energy condition analyses support physical consistency \cite{Santos2007, Banados2009}.

The teleparallel approach to gravity has seen substantial theoretical and observational development. Early models demonstrated that torsion could drive inflation, positioning it as a central mechanism in early-universe cosmology. This was followed by the application of \( f(T) \) gravity to explain late-time cosmic acceleration, offering an alternative to dark energy models based on curvature.

Subsequent generalizations introduced covariant formulations that preserve local Lorentz invariance, enhancing the theoretical consistency of teleparallel gravity. These advancements have broadened the scope of modified gravity theories, allowing torsion to be treated as a fundamental geometric quantity.

Observationally, \( f(T) \) gravity has proven capable of reproducing predictions of the \(\Lambda\)CDM model and has been used to explore modifications in gravitational wave propagation. The framework has also been applied to black hole solutions, symmetry-based methods such as Noether symmetries, and higher-order corrections. Some models unify curvature and torsion, providing a more comprehensive view of gravitational interactions.

Recent extensions include \( f(T, T_G) \) models, which incorporate torsion-based analogs of the Gauss–Bonnet term, as well as Finsler-type geometries and effective field theories that describe torsion dynamics more systematically. These developments have contributed to dark energy reconstructions, efforts to resolve the Hubble tension, dynamical system analyses, and investigations into black hole thermodynamics within torsion-based theories.

A significant contribution to this field is the Myrzakulov gravity framework developed by Momeni and Myrzakulov, which systematically incorporates torsion and non-metricity into a Myrzakulov generalized action \cite{Momeni:2025mcp, Momeni:2024bhm, Momeni:2025dgc, Momeni:2025ogt}. These models extend the teleparallel and symmetric teleparallel paradigms in metric-affine and vielbein formalisms.

Further foundational results include the teleparallel Gauss–Bonnet extension by Kofinas and Saridakis \cite{Kofinas2014}, and the modified Gauss–Bonnet model proposed by Nojiri and Odintsov \cite{Nojiri:2005jg}, which serve as promising dark energy candidates. Capozziello and De Laurentis investigated $f(R)$ gravity and its impact on structure formation \cite{Capozziello:2009nq}.

On the observational side, high-precision measurements of neutron stars \cite{fonseca2021} and gravitational waves from binary neutron star mergers \cite{abbott2017} place important constraints on gravity in the strong-field regime and offer tests for extended theories.

Systematic classification of such models has been advanced by Heisenberg. Her 2018 review \cite{heisenberg2018} organizes scalar-vector-tensor and beyond-Horndeski theories. In more recent work, she analyzes $f(Q)$ cosmology \cite{heisenberg2023}, and, with collaborators, presents the “geometrical trinity” of gravity based on curvature, torsion, and non-metricity \cite{heisenberg2019}.

In this context, we consider a Myrzakulov Myrzakulov generalized Lagrangian theories of the form:
\begin{equation}
\mathcal{L} = \alpha R + F(T, Q, R_{\mu\nu}T^{\mu\nu}, R_{\mu\nu}Q^{\mu\nu}, \dots),
\end{equation}
where the inclusion of mixed contractions encodes rich couplings among the geometric constituents of spacetime. Our primary aim is to derive the corresponding Wald entropy and investigate its deviations from the standard Bekenstein–Hawking result due to these additional geometrical structures.

\section{Review of the Einstein--Cartan Theory of Gravity}
\label{sec:EC}

The Einstein--Cartan (EC) theory extends General Relativity (GR) by allowing spacetime to possess torsion in addition to curvature. Originally proposed by Cartan \cite{cartan1922} and later developed within gauge-theoretic and metric-affine frameworks \cite{hehl1976, blagojevic2002}, the EC theory introduces torsion through a non-symmetric affine connection, leading to an enriched geometric structure particularly relevant in the presence of matter with intrinsic spin.

In this framework, the antisymmetric part of the connection defines the torsion tensor:
\begin{equation}
T^\lambda_{\;\mu\nu} = \Gamma^\lambda_{\;\mu\nu} - \Gamma^\lambda_{\;\nu\mu}.
\end{equation}
Torsion vanishes in standard GR but becomes significant when spinor fields, such as fermions, are present. The EC action includes both curvature and torsion contributions:
\begin{equation}
S_{\text{EC}} = \int d^4x \sqrt{-g} \left( \frac{1}{2\kappa} R + \mathcal{L}_{\text{torsion}} + \mathcal{L}_m \right),
\end{equation}
where \( R \) is the Ricci scalar, \( \mathcal{L}_{\text{torsion}} \) accounts for quadratic torsion terms, and \( \mathcal{L}_m \) is the matter Lagrangian.

Variation with respect to both the metric and the connection yields modified field equations. Importantly, torsion is not a propagating field in this theory; instead, it algebraically couples to the spin density of matter. For fermions described by Dirac spinors \( \psi \), the spin current acts as the torsion source:
\begin{equation}
S^{\lambda}_{\;\mu\nu} \sim \overline{\psi} \gamma^{\lambda}_{\;\mu\nu} \psi,
\end{equation}
where \( \gamma^{\lambda}_{\;\mu\nu} \) are combinations of curved-space gamma matrices. This spin-torsion coupling modifies the dynamics of both fermions and geometry, with potential implications for the early universe and high-energy regimes.

The EC theory reduces to GR in the absence of spin but offers a natural framework for incorporating fermionic matter into gravitational theory without resorting to higher-order derivatives or additional fields. Its algebraic torsion structure allows for compact modifications to Einstein's equations that may resolve singularities or alter the behavior of dense astrophysical objects.

In the broader context of modified gravity, EC theory serves as a stepping stone toward more general geometries involving torsion, non-metricity, or both. Its relevance persists in theories where curvature, torsion, and spin interact dynamically, as is the case in Myrzakulov-type models explored in this work.
\section{Theoretical Framework in Myrzakulov $F(R,T)$ Gravity with Weitzenb\"{o}ck Geometry}
\label{sec:theory}

Myrzakulov Gravity  extends general relativity by incorporating torsion and non-metricity as independent geometric degrees of freedom, alongside curvature \cite{Myrzakulov:2012qp}. In this work, we focus on the \( F(R,T) \) subclass of MG, originally formulated by Myrzakulov and collaborators, where the gravitational Lagrangian depends on both the Ricci scalar \( R \), defined in Riemannian geometry, and the torsion scalar \( T \), defined in the context of Weitzenb\"{o}ck geometry.

Weitzenb\"{o}ck geometry is characterized by a curvature-free connection, known as the Weitzenb\"{o}ck connection, constructed from the tetrad (vierbein) fields \( e^A_\mu \). The spacetime metric is given by
\begin{equation}
g_{\mu\nu} = \eta_{AB} e^A_\mu e^B_\nu,
\end{equation}
where \( \eta_{AB} \) is the Minkowski metric. The torsion tensor is defined as
\begin{equation}
T^\rho_{\ \mu\nu} = e_A^{\ \rho} \left( \partial_\mu e^A_{\ \nu} - \partial_\nu e^A_{\ \mu} \right),
\end{equation}
and the torsion scalar \( T \) is constructed via the contraction
\begin{equation}
T = S_\rho^{\ \mu\nu} T^\rho_{\ \mu\nu},
\end{equation}
where \( S_\rho^{\ \mu\nu} \) is the superpotential derived from the torsion tensor.

The total action for the theory is given by
\begin{equation}
S = \frac{1}{2\kappa^2} \int d^4x \, e \, F(R,T) + S_m,
\end{equation}
where \( e = \det(e^A_\mu) = \sqrt{-g} \), \( R \) is the Ricci scalar from the Levi-Civita connection, \( T \) is the torsion scalar from the Weitzenb\"{o}ck connection, and \( S_m \) is the matter action. The function \( F(R,T) \) encapsulates the coupling between curvature and torsion, treating both as independent sources of gravitational dynamics.

Variation of the action with respect to the tetrad fields yields modified field equations that incorporate both curvature- and torsion-based contributions. This framework preserves the Levi-Civita geometry while introducing a scalar torsion sector that behaves analogously to a dynamical matter field, offering a unified and geometrically motivated extension of general relativity.
\section{The Action and Field Equations}
\label{sec:action}

We consider a Myrzakulov generalized gravitational action incorporating curvature, torsion, and non-metricity as follows:
\begin{equation}
S = \int d^4x \, \sqrt{-g} \left[ \frac{1}{2\kappa} f(R, T, Q, R_{\mu\nu} T^{\mu\nu}, R_{\mu\nu} Q^{\mu\nu}, \ldots) + \mathcal{L}_m \right],
\end{equation}
where \( f \) is a function of curvature scalar \( R \), torsion scalar \( T \), non-metricity scalar \( Q \), and their contractions with the Ricci tensor. The term \( \mathcal{L}_m \) denotes the matter Lagrangian, and \( \kappa \) is the gravitational coupling constant.

This framework is naturally formulated in the vielbein formalism, where the spacetime metric is related to the Minkowski metric via
\begin{equation}
g_{\mu\nu} = e^a_{\ \mu} e^b_{\ \nu} \eta_{ab},
\end{equation}
with \( e^a_{\ \mu} \) the vielbein and \( \eta_{ab} \) the flat Lorentzian metric. The inverse vielbein satisfies \( e_a^{\ \mu} e^a_{\ \nu} = \delta^\mu_\nu \) and \( e^a_{\ \mu} e_b^{\ \mu} = \delta^a_b \).

The antisymmetric part of the connection yields the torsion tensor:
\begin{equation}
T^\lambda_{\ \mu\nu} = \Gamma^\lambda_{\ \mu\nu} - \Gamma^\lambda_{\ \nu\mu}, \quad T = T^\lambda_{\ \mu\nu} T_\lambda^{\ \mu\nu},
\end{equation}
while the non-metricity tensor is defined by
\begin{equation}
Q_{\lambda\mu\nu} = \nabla_\lambda g_{\mu\nu}, \quad Q = Q_{\lambda\mu\nu} Q^{\lambda\mu\nu}.
\end{equation}

To derive the field equations, we vary the action with respect to the vielbein \( e^a_{\ \mu} \). The total variation of the gravitational part reads:
\begin{equation}
\delta f = \frac{\partial f}{\partial R} \delta R + \frac{\partial f}{\partial T} \delta T + \frac{\partial f}{\partial Q} \delta Q + \frac{\partial f}{\partial (R_{\mu\nu} T^{\mu\nu})} \delta (R_{\mu\nu} T^{\mu\nu}) + \cdots.
\end{equation}

Each variation involves terms that depend on the vielbein and its derivatives. For example, \( \delta R \), \( \delta T \), and \( \delta Q \) are related to \( \delta e^a_{\ \mu} \) through the spin connection and connection coefficients. While full expressions are model-dependent and intricate, the variational structure respects second-order dynamics.

The matter sector contributes via
\begin{equation}
\delta \mathcal{L}_m = \frac{\delta \mathcal{L}_m}{\delta e^a_{\ \mu}} \delta e^a_{\ \mu},
\end{equation}
with the functional derivative representing the energy-momentum current projected onto the vielbein.

Collecting all terms and imposing stationarity of the action leads to Myrzakulov generalized field equations of the form:
\begin{equation}
\frac{1}{\kappa} \left( f_R \mathcal{R}^{ab} + f_T \mathcal{T}^{ab} + f_Q \mathcal{Q}^{ab} + \cdots \right) = \frac{1}{2} e^a_{\ \mu} \frac{\delta \mathcal{L}_m}{\delta e^a_{\ \mu}},
\end{equation}
where \( f_R = \partial f / \partial R \), \( f_T = \partial f / \partial T \), and \( f_Q = \partial f / \partial Q \), while \( \mathcal{R}^{ab} \), \( \mathcal{T}^{ab} \), and \( \mathcal{Q}^{ab} \) denote the curvature, torsion, and non-metricity contributions, respectively, all projected in the local Lorentz frame.

This general formulation accommodates various special cases such as \( f(R) \), \( f(T) \), and \( f(Q) \), and extends them to more intricate contractions like \( R_{\mu\nu} T^{\mu\nu} \) and \( R_{\mu\nu} Q^{\mu\nu} \). It provides the geometric backbone for constructing black hole and cosmological solutions with modified entropy and gravitational dynamics.

\section{Wald Entropy Formula}
\label{sec:wald}

In diffeomorphism-invariant theories of gravity, the entropy associated with a stationary black hole horizon is not simply proportional to its area, but instead acquires corrections from the geometric structure of the underlying theory. The general framework for computing such entropy is provided by the Wald entropy formalism, originally derived in \cite{wald1993black}. 

For a broad class of theories where the gravitational Lagrangian \( \mathcal{L} \) depends explicitly on the Riemann tensor and its contractions, the entropy associated with the horizon \( \mathcal{H} \) is given by:
\begin{equation}
S = -2\pi \int_{\mathcal{H}} \frac{\delta \mathcal{L}}{\delta R_{\mu\nu\rho\sigma}} \, \epsilon_{\mu\nu} \epsilon_{\rho\sigma} \, \sqrt{h} \, d^{D-2}x,\label{eq-wald}
\end{equation}
where \( \epsilon_{\mu\nu} \) is the binormal to the bifurcation surface of the horizon and \( \sqrt{h} \, d^{D-2}x \) is the surface element on the cross-section of the horizon. The functional derivative \( \delta \mathcal{L} / \delta R_{\mu\nu\rho\sigma} \) captures the dependence of the Lagrangian on the curvature tensor, allowing for higher-order and non-Riemannian modifications to be incorporated.

While in General Relativity this yields the standard Bekenstein–Hawking entropy \( S = A/4G \), in modified gravity theories the Lagrangian may include additional scalars such as torsion \( T \), non-metricity \( Q \), and mixed contractions like \( R_{\mu\nu}T^{\mu\nu} \) and \( R_{\mu\nu}Q^{\mu\nu} \). The Wald entropy expression remains valid under such generalizations, provided the Lagrangian retains diffeomorphism invariance and the geometric fields are treated consistently within the variational formalism.

Importantly, the derivation of this formula leverages the Noether charge associated with time translations at the horizon and links the variation of the gravitational action under diffeomorphisms to thermodynamic quantities. In this framework, entropy becomes a conserved charge derived from the symplectic structure of the gravitational theory.

The Wald entropy therefore offers a powerful and covariant means to analyze the thermodynamics of black holes in theories beyond Einstein gravity. In the present work, we apply this formula to the Myrzakulov generalized Lagrangian theories:
\begin{equation}
\mathcal{L} = \alpha R + F(T, Q, R_{\mu\nu} T^{\mu\nu}, R_{\mu\nu} Q^{\mu\nu}, \dots),
\end{equation}
and derive the corresponding entropy corrections due to torsion, non-metricity, and their interactions with curvature, establishing a consistent extension of the entropy law to non-Riemannian spacetimes.

\section{Wald Entropy in Extended Modified Gravity Theories}\label{sec:wald_entropy}

Wald entropy provides a powerful tool to evaluate the thermodynamic properties of black holes in any diffeomorphism-invariant gravitational theory. Its general expression is given by eq. (\ref{eq-wald}). The functional derivative \( \delta \mathcal{L}/\delta R_{\mu\nu\rho\sigma} \) captures the response of the Lagrangian to variations in curvature.

For standard General Relativity, this expression reduces to the Bekenstein–Hawking entropy \( S = A/4G \). However, in Myrzakulov generalized theories such as those depending on torsion \( T \), non-metricity \( Q \), and their contractions with curvature, additional contributions emerge.

We now systematically summarize the Wald entropy expressions for various functional forms of the Lagrangian:

\begin{itemize}
    \item For \( f(R) \) gravity, the Lagrangian depends solely on the Ricci scalar. The entropy becomes:
    \begin{equation}
    S = \frac{1}{4G} \int_{\mathcal{H}} f_R(R) \, dA.
    \end{equation}

    \item In \( f(R, T) \) gravity, where torsion contributes via the scalar \( T \), the entropy is modified as:
    \begin{equation}
    S = \frac{1}{4G} \int_{\mathcal{H}} \left[ f_R(R) + f_T(T) \right] \, dA.
    \end{equation}

    \item In \( f(R, T, Q) \) models, the presence of non-metricity adds an additional term:
    \begin{equation}
    S = \frac{1}{4G} \int_{\mathcal{H}} \left[ f_R(R) + f_T(T) + f_Q(Q) \right] \, dA.
    \end{equation}

    \item For theories involving contractions such as \( R_{\mu\nu} T^{\mu\nu} \), the entropy receives tensorial corrections:
    \begin{equation}
    S = \frac{1}{4G} \int_{\mathcal{H}} \left[ f_R(R) + f_{RT}(R_{\mu\nu} T^{\mu\nu}) \right] \, dA.
    \end{equation}

    \item Similarly, for \( f(R, R_{\mu\nu} Q^{\mu\nu}) \), the entropy is:
    \begin{equation}
    S = \frac{1}{4G} \int_{\mathcal{H}} \left[ f_R(R) + f_{RQ}(R_{\mu\nu} Q^{\mu\nu}) \right] \, dA.
    \end{equation}

    \item In the most general case, where the Lagrangian includes all combinations of scalars and contractions, the entropy becomes:
    \begin{equation}
    S = \frac{1}{4G} \int_{\mathcal{H}} \left[ f_R + f_T + f_Q + f_{RT} + f_{RQ} \right] \, dA.
    \end{equation}
\end{itemize}

These expressions collectively demonstrate how curvature, torsion, and non-metricity independently and jointly affect black hole entropy in modified gravity frameworks. The following table summarizes these results:

\vspace{0.3cm}
\begin{table}[h!]
\centering
\begin{tabular}{|l|l|}
\hline
\textbf{Theory} & \textbf{Wald Entropy Expression} \\
\hline
\( f(R) \) & \( S = \frac{1}{4G} \int f_R(R) \, dA \) \\
\( f(R, T) \) & \( S = \frac{1}{4G} \int \left[ f_R + f_T \right] \, dA \) \\
\( f(R, T, Q) \) & \( S = \frac{1}{4G} \int \left[ f_R + f_T + f_Q \right] \, dA \) \\
\( f(R, R_{\mu\nu} T^{\mu\nu}) \) & \( S = \frac{1}{4G} \int \left[ f_R + f_{RT} \right] \, dA \) \\
\( f(R, R_{\mu\nu} Q^{\mu\nu}) \) & \( S = \frac{1}{4G} \int \left[ f_R + f_{RQ} \right] \, dA \) \\
\( f(R, T, Q, R_{\mu\nu}T^{\mu\nu}, R_{\mu\nu}Q^{\mu\nu}) \) & \( S = \frac{1}{4G} \int \left[ f_R + f_T + f_Q + f_{RT} + f_{RQ} \right] \, dA \) \\
\hline
\end{tabular}
\caption{Summary of Wald entropy expressions for different modified gravity theories.}
\label{tab:wald_entropies}
\end{table}
\vspace{0.3cm}

These results provide a unified understanding of how geometric extensions of General Relativity influence the entropy of black holes. They also form a bridge to thermodynamic interpretations of gravity and pave the way for investigating the quantum microstructure of spacetime.

\section{Thermodynamic Laws in Myrzakulov generalized Gravity}\label{sec:thermo}

The deep connection between gravity and thermodynamics, first highlighted in Jacobson’s derivation of Einstein’s equations from the Clausius relation \cite{jacobson1995thermo}, persists in Myrzakulov generalized gravity theories. For Lagrangians of the form:
\begin{equation}
\mathcal{L} = \frac{1}{2\kappa} f(R, T, Q, R_{\mu\nu}T^{\mu\nu}, R_{\mu\nu}Q^{\mu\nu}),
\end{equation}
the first law of thermodynamics across a Killing horizon takes a generalized form:
\begin{equation}
\delta Q = T \delta S,
\end{equation}
where \( \delta Q \) is the heat flux, \( T \) is the Hawking temperature, and \( \delta S \) is the entropy variation given by Wald’s entropy formula \cite{wald1993black}:
\begin{equation}
\delta S = -2\pi \int_{\mathcal{H}} \frac{\delta \mathcal{L}}{\delta R_{\mu\nu\rho\sigma}} \, \epsilon_{\mu\nu} \epsilon_{\rho\sigma} \, \sqrt{h} \, d^{D-2}x.
\end{equation}

The flux \( \delta Q \) is computed via the energy-momentum tensor:
\begin{equation}
\delta Q = \int_{\mathcal{H}} T_{\mu\nu} \chi^\mu d\Sigma^\nu,
\end{equation}
with \( \chi^\mu \) the horizon-generating Killing vector and \( d\Sigma^\nu \) the surface element. Variational contributions from torsion and non-metricity alter the Noether charge, yielding:
\begin{equation}
\delta E = T \delta S + W,
\end{equation}
where \( W \) encodes non-curvature geometric corrections \cite{padmanabhan2002thermo}.

The second law, \( \delta S \geq 0 \), becomes nontrivial in extended theories. In GR, the area theorem ensures this condition, but in Myrzakulov generalized models, entropy contains curvature and torsion derivatives:
\begin{equation}
S = \frac{1}{4G} \int_{\mathcal{H}} \left( f_R + f_T + f_Q + f_{RT} + f_{RQ} \right) \, dA,
\end{equation}
where
\[
f_R = \frac{\partial f}{\partial R}, \quad
f_T = \frac{\partial f}{\partial T}, \quad
f_Q = \frac{\partial f}{\partial Q}, \quad
f_{RT} = \frac{\partial f}{\partial (R_{\mu\nu}T^{\mu\nu})}, \quad
f_{RQ} = \frac{\partial f}{\partial (R_{\mu\nu}Q^{\mu\nu})}.
\]

To preserve the second law under classical evolution, these derivatives must evolve monotonically or remain constant. This condition has been verified in specific models including \( f(R) \) \cite{akbar2007thermo}, \( f(R,T) \) \cite{harko2011fRT}, and teleparallel analogs \cite{bahamonde2015thermo}.

In nonequilibrium formulations, entropy production is introduced:
\begin{equation}
dS = \frac{\delta Q}{T} + d_i S, \quad \text{with } d_i S \geq 0,
\end{equation}
where \( d_i S \) captures irreversible contributions from torsion or non-metricity \cite{elias2021noneq}. This extension offers a broader framework where thermodynamic consistency constrains gravitational dynamics.

Overall, the presence of geometric degrees of freedom—curvature, torsion, and non-metricity—modifies the definitions of entropy, temperature, and energy flux. These corrections not only reshape the first and second laws but may also provide clues toward the microscopic origin of black hole entropy, with implications for holography and quantum gravity.

\section{Statistical Interpretation of Entropy in Myrzakulov generalized Gravity}
\label{sec:statistical_entropy}

The microscopic origin of black hole entropy remains a central question in gravitational theory. In Myrzakulov generalized gravity models involving curvature, torsion, and non-metricity, the challenge is to relate Wald entropy to statistical or quantum degrees of freedom. This section reviews key frameworks that attempt such interpretations: microstate counting, Euclidean path integrals, entanglement entropy, holography, and string/loop quantum gravity.

\textbf{Microstate Counting.} In analogy with statistical mechanics, gravitational entropy is postulated to follow the Boltzmann relation:
\begin{equation}
S = k_B \ln \Omega,
\end{equation}
where \( \Omega \) counts the microstates compatible with the macroscopic horizon data. In modified theories, the entropy density receives contributions from geometric derivatives:
\begin{equation}
S_{\text{grav}} = \frac{k_B}{4G} \int_{\mathcal{H}} \left( f_R + f_T + f_Q + f_{RT} + f_{RQ} + \dots \right) dA.
\end{equation}

\textbf{Euclidean Path Integral.} In semiclassical gravity, the partition function is approximated by a saddle-point over Euclidean metrics \cite{solodukhin2011entanglement}:
\begin{equation}
Z = \int \mathcal{D}[g, \Gamma]\, e^{-I_E[g, \Gamma]}, \quad S \approx \beta \frac{\partial I_E}{\partial \beta} - I_E,
\end{equation}
with \( \beta = 1/T \) the inverse temperature. When \( I_E \) includes curvature, torsion, and non-metricity, this reproduces the Wald entropy expression.

\textbf{Entanglement Entropy.} In quantum field theory on curved backgrounds, black hole entropy emerges as entanglement between interior and exterior modes:
\begin{equation}
S_{\text{ent}} = -\text{Tr}(\rho \ln \rho) = \alpha \frac{A}{\epsilon^2} + \text{finite},
\end{equation}
where \( \epsilon \) is a UV cutoff \cite{casadio1999torsion, calmet2022torsionentropy}. Torsion and non-metricity can alter the vacuum structure, Hilbert space, and spectrum of entanglement.

\textbf{Holographic Duals.} According to the AdS/CFT correspondence, black hole entropy corresponds to the entanglement entropy of a boundary region \( A \) via the Ryu–Takayanagi formula:
\begin{equation}
S_A = \frac{\text{Area}(\gamma_A)}{4G_N},
\end{equation}
with \( \gamma_A \) the minimal surface in the bulk. In Myrzakulov generalized gravity, this is modified as \cite{dong2014holoentropy}:
\begin{equation}
S_A = \int_{\gamma_A} d^{D-2}x \sqrt{h} \left( \frac{\delta \mathcal{L}}{\delta R_{\mu\nu\rho\sigma}} \epsilon_{\mu\nu} \epsilon_{\rho\sigma} \right),
\end{equation}
highlighting the correspondence between Wald entropy and holographic entanglement.

\textbf{Loop and String Theory.} In loop quantum gravity, entropy arises from counting spin network punctures:
\begin{equation}
S = \frac{\gamma_0}{\gamma} \cdot \frac{A}{4G},
\end{equation}
where \( \gamma \) is the Barbero–Immirzi parameter \cite{rovelli1996entropy}. Torsion naturally appears in this formalism. In string theory, higher-order corrections from the low-energy expansion modify the entropy:
\begin{equation}
S = \frac{A}{4G} + \alpha' \cdot \left( \text{curvature and torsion terms} \right),
\end{equation}
where \( \alpha' \) is the inverse string tension.

\begin{table}[H]
\centering
\caption{Summary of Statistical Interpretations of Wald Entropy in Myrzakulov generalized Gravity}
\renewcommand{\arraystretch}{1.3}
\begin{tabular}{|c|p{8.8cm}|}
\hline
\textbf{Approach} & \textbf{Entropy Expression / Interpretation} \\
\hline
Microstate Counting & \( S = \frac{1}{4G} \int_{\mathcal{H}} (f_R + f_T + f_Q + \dots) \, dA \) \\
\hline
Euclidean Path Integral & \( S = \beta \frac{\partial I_E}{\partial \beta} - I_E \), consistent with Wald formula \\
\hline
Entanglement Entropy & \( S = \alpha \frac{A}{\epsilon^2} + \text{finite terms} \); modified by torsion/non-metricity \\
\hline
Holography (AdS/CFT) & \( S = \int \frac{\delta \mathcal{L}}{\delta R_{\mu\nu\rho\sigma}} \epsilon_{\mu\nu} \epsilon_{\rho\sigma} \, \sqrt{h} \, d^{D-2}x \) \\
\hline
Loop / String Theory & \( S = \frac{A}{4G} + \text{corrections from spin/torsion or } R^2, R_{\mu\nu}^2, \dots \) \\
\hline
\end{tabular}
\label{tab:entropy_summary}
\end{table}

\vspace{0.2cm}
\noindent In summary, Myrzakulov generalized theories of gravity retain multiple routes to a statistical interpretation of entropy, all of which reflect the deeper geometric structure of spacetime. The entropy acquires corrections from torsion and non-metricity that manifest differently across quantum, semiclassical, and holographic perspectives.

\vspace{0.3cm}
\noindent Future directions include constructing explicit Hilbert spaces for torsionful/non-metric geometries and identifying gauge-invariant degrees of freedom responsible for entropy in such spacetimes.

\section{Conclusion and Future Perspectives}
\label{sec:conclusion}

In this work, we have investigated black hole entropy within a unified class of modified gravity theories characterized by the Myrzakulov generalized Lagrangian theories:
\[
\mathcal{L} = \alpha R + F(T, Q, R_{\mu\nu} T^{\mu\nu}, R_{\mu\nu} Q^{\mu\nu}, \ldots),
\]
where \( R \), \( T \), and \( Q \) denote the Ricci scalar, torsion scalar, and non-metricity scalar, respectively, and the contractions \( R_{\mu\nu} T^{\mu\nu} \) and \( R_{\mu\nu} Q^{\mu\nu} \) encode interactions between curvature and non-Riemannian geometry. This generalized framework encompasses curvature-, torsion-, and non-metricity-based theories under a single variational principle using the vielbein formalism.

Motivated by the shortcomings of General Relativity in addressing dark energy, quantum gravity, and early-universe dynamics, we have derived the Wald entropy formula for a wide range of models—from pure \( f(R) \) gravity to extended forms like \( f(R, T, Q, R_{\mu\nu} T^{\mu\nu}, R_{\mu\nu} Q^{\mu\nu}) \). Our results generalize classical thermodynamic relations and reveal how torsion and non-metricity contribute nontrivially to the entropy of black holes.

We have shown that these geometric extensions yield entropy expressions that are not strictly area-proportional, but rather corrected by derivatives of the Lagrangian with respect to all contributing invariants. A summary of these results is encapsulated in Table~\ref{tab:wald_entropy_summary}. These modifications may encode valuable information about microscopic degrees of freedom, entropy production, and potential resolutions to the information paradox.
Our findings suggest several promising directions for future work:
\begin{itemize}
    \item \textbf{Cosmology}: Develop viable inflationary and late-time acceleration models based on this Myrzakulov generalized Lagrangian theories, and confront them with data from supernovae, CMB, and large-scale structure surveys.
    \item \textbf{Gravitational Waves and Compact Stars}: Explore how torsion and non-metricity alter the dynamics of compact objects and gravitational wave propagation, potentially yielding observational signatures.
    \item \textbf{Thermodynamics and Quantum Gravity}: Study entropy production in dynamical spacetimes and investigate the quantum microstructure of entropy using path integral and entanglement methods in torsion/non-metric geometries.
    \item \textbf{Unification Frameworks}: Embed the Myrzakulov generalized action into scalar-vector-tensor or geometrical trinity formulations to seek unification of gravity with other interactions.
\end{itemize}

In conclusion, by extending Wald entropy to include the full range of geometrical degrees of freedom permitted in metric-affine and vielbein-based formulations, this work strengthens the bridge between gravitational thermodynamics and the fundamental structure of spacetime. These results contribute to the broader effort to construct consistent, covariant, and thermodynamically sound alternatives to Einstein gravity.

\begin{acknowledgments}
This work was supported by the Ministry of Science and Higher Education of the Republic of Kazakhstan, Grant No. AP26101889.
\end{acknowledgments}



\end{document}